\documentclass[prl,twocolumn,amsmath,amssymb]{revtex4}

\usepackage[dvips]{graphicx}
\usepackage{graphics}
\usepackage{dcolumn}
\usepackage{bm}
\usepackage[usenames]{color}

\begin{document}

\title{\begin{center} Transport Characterization of the Magnetic Anisotropy of (Ga,Mn)As
 \end{center}}

\author{K. Pappert}
\affiliation{Physikalisches Institut (EP3), Universit\"{a}t
W\"{u}rzburg, Am Hubland, D-97074 W\"{u}rzburg, Germany}

\author{S. H\"{u}mpfner}
\affiliation{Physikalisches Institut (EP3), Universit\"{a}t
W\"{u}rzburg, Am Hubland, D-97074 W\"{u}rzburg, Germany}

\author{J. Wenisch}
\affiliation{Physikalisches Institut (EP3), Universit\"{a}t
W\"{u}rzburg, Am Hubland, D-97074 W\"{u}rzburg, Germany}

\author{K. Brunner}
\affiliation{Physikalisches Institut (EP3), Universit\"{a}t
W\"{u}rzburg, Am Hubland, D-97074 W\"{u}rzburg, Germany}

\author{C. Gould}
\affiliation{Physikalisches Institut (EP3), Universit\"{a}t W\"{u}rzburg, Am Hubland, D-97074 W\"{u}rzburg, Germany}

\author{G. Schmidt}
\affiliation{Physikalisches Institut (EP3), Universit\"{a}t W\"{u}rzburg, Am Hubland, D-97074 W\"{u}rzburg, Germany}

\author{L.W. Molenkamp}
\affiliation{Physikalisches Institut (EP3), Universit\"{a}t W\"{u}rzburg, Am Hubland, D-97074 W\"{u}rzburg, Germany}

\date{\today}

\begin{abstract}

The rich magnetic anisotropy of compressively strained (Ga,Mn)As
has attracted great interest recently. Here we discuss a sensitive
method to visualize and quantify the individual components of the
magnetic anisotropy using transport. A set of high resolution
transport measurements is compiled into color coded resistance
polar plots, which constitute a fingerprint of the symmetry
components of the anisotropy. As a demonstration of the
sensitivity of the method, we show that these typically reveal the
presence of both the $[\overline{1}10]$ and the [010] uniaxial
magnetic anisotropy component in (Ga,Mn)As layers, even when most
other techniques reveal only one of these components.

\end{abstract}


\maketitle

The ferromagnetic semiconductor (Ga,Mn)As exhibits rich magnetic
anisotropy behaviour. Experimental studies of compressively
strained (Ga,Mn)As based on direct magnetization measurements by
SQUID magnetometry \cite{Sawicki} show a principally biaxial
anisotropy in the sample plane for highly doped samples at 4 K.
Additionally, uniaxial anisotropy components along
$[\overline{1}10]$ \cite{Roukes} or [010] \cite{ChrisTAMR} have
been reported. The $[\overline{1}10]$ uniaxial anisotropy is
widely accepted to be present in most (Ga,Mn)As layers as a
secondary component at 4 K which gets stronger upon annealing, or
increasing sample temperature. The primary biaxial anisotropy is
expected from theory \cite{theory1,theory2} and originates from
the hole-mediated ferromagnetism and the strong spin orbit
coupling, which links the magnetic interactions to the lattice
structure. This theory also provides a partial understanding of
the emergence of a different easy axis for higher carrier
concentration or temperature, in that it predicts a transition
from biaxial along $\langle100\rangle$ to biaxial along
$\langle110\rangle$. It does not however account for the presence
of either uniaxial term. Nevertheless, the existence of these
uniaxial easy axes is experimentally well established, and it
remains of great interest, both as a practical matter in designing
and controlling magnetization in experimental samples and as a
fundamental tool for increasing our understanding of the
magnetization behaviour in strong spin-orbit systems, to develop
tools to precisely investigate the magnetic anisotropy in
(Ga,Mn)As.

Traditionally the anisotropy in these materials has been
investigated by direct measurement of the projection of the
magnetization onto characteristic directions using SQUID or VSM.
The advent of vector field magnets has recently opened up new
possibilities for acquiring a detailed mapping of the anisotropy.
We present one such method here, based on summarizing the results
of transport measurements into color coded resistance polar plots
(RPP) which act as fingerprints for the anisotropy of a given
structure. Not only is this method faster than the traditional
alternatives, but it is also more sensitive to certain secondary
components of the anisotropy, in particular those with easy axes
collinear to the primary anisotropy\cite{Cowburn}. The technique
thus often reveals the existence of components which would be
missed using standard techniques. Moreover, the technique can be
applied to study the anisotropy of layers by using macroscopic
transport structures, or applied directly to device elements. It
can thus reveal any impacts of processing or the influence of
small strain fields due to, for example, contacting.

\begin{figure}[h!]
\includegraphics[angle=0,width=7 cm]{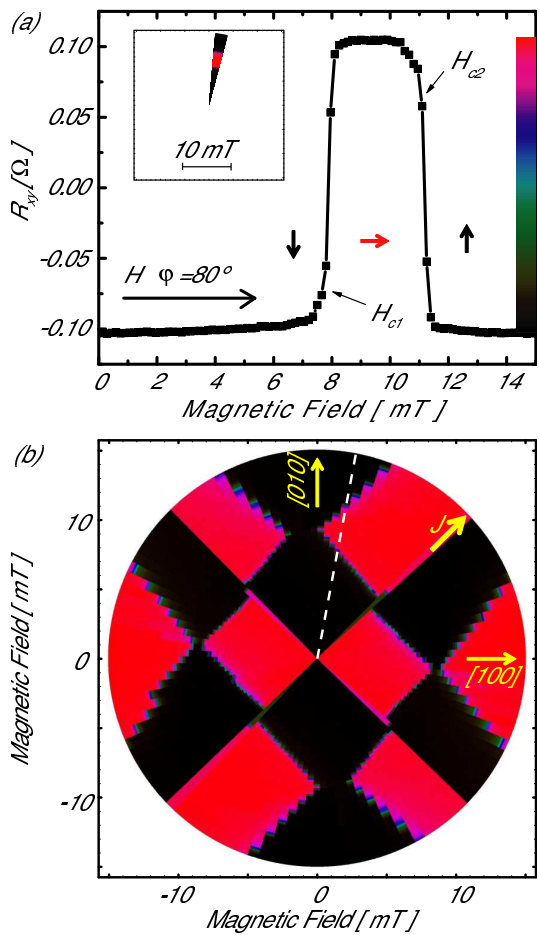}
\vspace*{0.0cm} \caption{(a) In-plane Hall measurement along
$\varphi=80^{\circ}$ with marked first and second switching field,
color scale and the corresponding section of a color coded
resistance polar plot(inset). (b) Resistance polar plot from a
full set of In-plane Hall measurements along every $3^\circ$. The
$80^\circ$-section corresponding to (a) is marked by a dashed
line.} \label{figure1} \vspace*{-0.2cm}
\end{figure}

Here, we apply the RPP technique to verify that both the [010] and
the $[\overline{1}10]$ uniaxial anisotropy terms can be, and
typically are, simultaneously present in compressively strained
(Ga,Mn)As layers at 4 K. Color coded polar plots can be compiled
from any measurement data that shows a response to the
magnetization direction. Anisotropic
magnetoresistance(AMR)\cite{Jan}, tunneling anisotropic
magnetoresistance (TAMR)\cite{ChrisTAMR} or in-plane Hall(IPH)
measurements\cite{Roukes} are typical examples in (Ga,Mn)As. For
example, sending a current \textbf{J} through a (Ga,Mn)As Hall bar
device along $\vartheta=0^\circ$, yields a
$\sin(2\vartheta)$-dependence of the transverse resistance
$R_{xy}$ on the magnetization direction $\vartheta$ due to the IPH
effect\cite{Jan}: $R_{xy}(\vartheta)=(\Delta R_{xy})
\sin(2\vartheta)\label{eq1}$

We demonstrate the usefulness of the technique by applying it to
the characterization of a typical 20 nm thick (Ga,Mn)As layer
grown on a GaAs (001) substrate by low-temperature molecular beam
epitaxy. It is patterned into a 60 $\mu$m wide Hall bar oriented
along the $\langle110\rangle$ crystal direction by optical
lithography and chlorine assisted dry etching. Contacts are
established through metal evaporation and lift off.

Transport measurements are carried out in a magnetocryostat fitted
with a vector field magnet comprised of three mutually orthogonal
pairs of Helmholtz coils that allow the application of a magnetic
field of up to 300 mT in any direction. An angular set of IPH
curves is acquired while sweeping the magnetic field along
multiple directions in the sample plane. For each individual angle
the magnetization state of the sample is first prepared by a
strong negative magnetic field along $\varphi$. The field is then
slowly brought down to zero while assuring that the field vector
never deviates from the $-\varphi$ direction. The IPH curve as a
function of positive magnetic field in the $\varphi$ direction is
then acquired from zero to higher fields, and these results are
displayed in a RPP as in Fig.~\ref{figure1}b.

Fig.~\ref{figure1}a shows a typical IPH measurement. After
magnetizing the sample along $80^\circ$ at -300 mT, the
magnetization relaxes towards the $[0\overline{1}0]$ easy axis, as
the field is brought back to zero. This corresponds to the
negative resistance state associated with an angle of $\vartheta=
-135^\circ$ between \textbf{J} and \textbf{M}. The first abrupt
resistance change at the field $H_{c1}$ corresponds to a
$90^\circ$ reorientation of \textbf{M} towards the other (Ga,Mn)As
easy axis $\sim[001]$.  A second reorientation of \textbf{M}
towards [010] at $H_{c2}$ completes the magnetisation reversal. A
set of such magnetic field scans along many angles, here every
$3^\circ$, is compiled into a RPP with the magnetic field $H$
along the radius and each scan at its angle $\varphi$.

The inset of Fig.~\ref{figure1}a shows the $80^\circ$-segment of
the full RPP of Fig.~\ref{figure1}b. The intensity encodes the
normalized resistance value, where low and high denote the minimum
and maximum resistance of the entire curve set, respectively. The
positions of the switching fields in the polar plot and the
symmetry of the pattern they form contain information on the
underlying magnetic anisotropy.

We now explain the observed pattern and the conclusions that we may
infer from it. As already noted, (Ga,Mn)As layers grown on a GaAs
(001) substrate are compressively strained and exhibit a biaxial
magnetic anisotropy ($K_{1}$) with the two easy axes along [100] and
[010] at 4 K. Additional smaller uniaxial anisotropy components
collinear to one of the biaxial easy axes, for example [010]
($K_{u1}$) \cite{ChrisTAMR} or bisecting the biaxial easy axes,
along $[\overline{1}10]$ ($K_{u2}$)\cite{Roukes} have been
previously reported. The energy $E$ of a single (Ga,Mn)As domain
with magnetization $M$ pointing in $\vartheta$- and external
magnetic field $H$ in $\varphi$-direction is thus phenomenologically
expressed as:

\vspace*{-.2 cm}
\begin{equation}\label{eq2} 
E = \frac{K_{1}}{4}\sin^{2}(2 \vartheta)+K_{u1}\sin^{2}(\vartheta)
+ K_{u2}\sin^{2}(\vartheta-45^\circ)-MH\cos(\vartheta-\varphi)
\end{equation}

where the last term is the Zeeman energy. When an external magnetic
field is applied, the magnetization direction follows a local energy
minimum (coherent or Stoner-Wohlfarth rotation). However, if the
energy gained by a magnetization reorientation to any other energy
minimum is larger than the respective domain wall (DW)
nucleation/propagation energy $\varepsilon$, a DW is nucleated and
propagates through the structure, resulting in an abrupt
magnetization reversal. Due to the mainly biaxial nature of the
magnetic anisotropy the magnetization typically reverses in a
double-step switching process through the nucleation and propagation
of two $90^\circ$ DW, as in Fig.~\ref{figure1}a.

\begin{figure}[h!]
\includegraphics[angle=0,width=8cm]{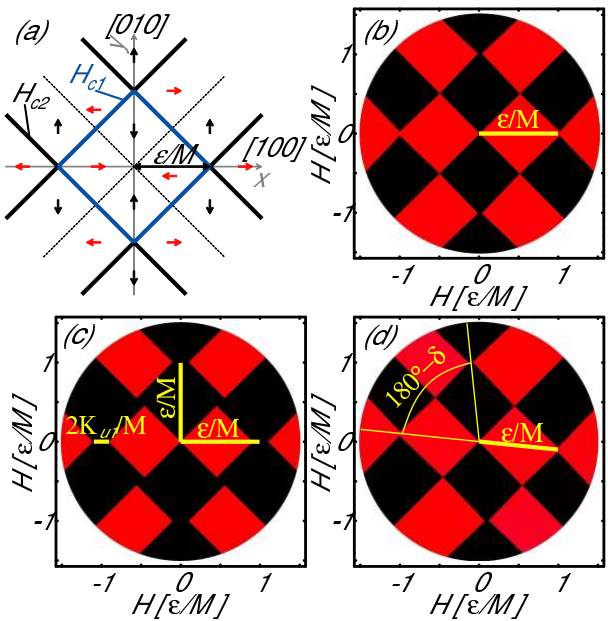}
\vspace*{0.0cm} \caption{(a) Switching field positions (thick
solid lines) in a polar plot for a biaxial material with easy axes
along [100] and [010](gray). The magnetization direction in each
region of the plot is indicated by arrows (red/black: high/low
resistance) and the hard axes by dashed lines. (b-d) Simulated
resistance polar plots for a biaxial material with easy axes along
the [100] and [010] crystal directions(b) and the same material
with an additional uniaxial anisotropy along [010](c) or
$[1\overline{1}0]$(d). Color scale of the resistance as in
Fig.~\ref{figure1}. $\varepsilon$ denotes the $90^\circ$-DW
nucleation/propagation energy.} \label{figure2} \vspace*{0.0cm}
\end{figure}

We first discuss the regime, where DW nucleation/propagation
dominates the magnetization reversal process, i.e. where
$\varepsilon$ is much smaller than the crystalline anisotropy. We
start with the pure biaxial magnetic anisotropy, i.e., with
$K_{u1}=K_{u2}=0$. During the double-step switching process, the
magnetostatic energy minima remain to a good approximation along
the biaxial easy axes, whose difference in energy is given by the
Zeeman term in Eq. \ref{eq2}. When the energy gained through a
$90^\circ$ magnetization reorientation is larger than
$\varepsilon_{90^\circ}$, the nucleation and propagation energy of
a $90^\circ$-DW, a thermally activated switching event becomes
possible, which on the timescale of our measurement, results in an
immediate switching event. The polar plot in Fig.~\ref{figure2}a
shows the characteristic square pattern formed by these switching
fields \cite{Cowburn}, Fig.~\ref{figure2}b the corresponding
calculated RPP. The switching field positions (thick black lines)
can be expressed in cartesian coordinates ($x=H_{c} \cos\varphi;
y=H_{c} \sin\varphi$) as given in Tab.~\ref{tab1}. The diagonals
(grey) of the $H_{c1}$-square represent the biaxial easy axes of
the material(here along $0^\circ$ and $90^\circ$). The diagonal's
length is equal to $\frac{2\varepsilon_{90^\circ}}{M}$. The dashed
lines represent the hard magnetic axes. The arrows illustrate the
direction of the magnetization and their color the corresponding
resistance state of the respective section.

\begin{table}[h!]
\begin{tabular}{|c|c|}
\hline Anisotropy components & $H_{c}$-positions in polar plot\\
\hline
 & \\
$\langle 100 \rangle$& $y=\pm x \pm \frac{\varepsilon}{M}$\\
 & \\
$\langle100\rangle$ + $[010]$& $y=\pm x \pm\frac{\varepsilon}{M}\pm\frac{K_{u1}}{M} $\\
 & \\
$\langle100\rangle$ + $[\overline{1}10]$&

$y=
\begin{cases} + x \pm \sqrt{2}
\frac{\varepsilon}{M}\cos(45^\circ-\delta/2)
\\
- x \pm \sqrt{2} \frac{\varepsilon}{M}\cos(45^\circ+\delta/2)
\end{cases}$\\

\hline
\end{tabular}
\caption{Positions of the switching fields in a polar plot given
in cartesian coordinates ($x=H_{c} \cos\varphi; y=H_{c}
\sin\varphi$). $\delta/2$ is the angle between global and biaxial
easy axis as in Fig.~\ref{figure2}d(see text).}\label{tab1}
\vspace*{-0.5cm}
\end{table}

An additional small uniaxial anisotropy $K_{u1}$ along one of the
biaxial easy axes (here along $90^\circ$) alters this pattern as
shown in Fig.~\ref{figure2}c. The symmetry between the two biaxial
easy axes is lifted, since one of them is parallel ($B_eU_e$)  and
one perpendicular ($B_eU_h$) to the easy axis of the uniaxial
component. The switching field positions in Tab. \ref{tab1} can be
derived as discussed above\cite{Cowburn}. $90^\circ$-switches away
from (towards) the $B_eU_e$ axis occur now at higher (lower)
magnetic fields as compared to the pure biaxial anisotropy.
Typical steps in the $H_{c1}$- pattern emerge
(Fig.~\ref{figure2}c, e.g. along $45^\circ$). The strength of the
uniaxial anisotropy can be determined from the separation $\frac{
2K_{u1} }{M}$ between $H_{c1}$ and $H_{c2}$ along the $B_eU_h$
axis. Another characteristic feature is the "open corner" of the
$H_{c1}$- pattern along the $B_eU_e$ axis. A $180^\circ$
magnetization reversal becomes energetically more favorable in
this angular region, than two successive $90^\circ$ events
\cite{DWs,Cowburn}. Since the isotropic magnetoresistance of
typical samples is relatively small, two magnetization directions
differing by $180^\circ$ are not distinguishable on the scale
considered here, and have the same color in the RPP.

If on the other hand, apart from the main biaxial anisotropy
($K_{1}$,$\langle100\rangle$), a uniaxial anisotropy $K_{u2}$ with
easy axis along $[\overline{1}10]$ (here $135^\circ$) is also
present and assuming that the DW nucleation/propagation energy
depends on the angle $\Delta\vartheta$ between the two domains as
$\varepsilon_{\Delta\vartheta}=\varepsilon_{90^\circ}(1-\cos(\Delta\vartheta))$,
the calculation gives the RPP of Fig.~\ref{figure2}d. The uniaxial
anisotropy component flattens the energy surface and shifts the
positions of the biaxial energy minima by
$\frac{\delta}{2}=\frac{1}{2}\arcsin(\frac{K_{u2}}{K_{1}})$
towards the uniaxial easy axis\cite{Bland}. The resulting "global
easy axes" appear as diagonals of a characteristic rectangle in
the RPP, where the uniaxial easy axis is along the median line of
the longer edge. The positions of the switching events
(Tab.~\ref{tab1}) can be derived equating $\varepsilon$ with the
difference in Zeeman energy between the minima. The diagonal's
length is again equal to $\frac{2\varepsilon_{90^\circ}}{M}$.

We demonstrate the usefulness of the technique by applying it to
the characterization of the typical (Ga,Mn)As sample discusses in
Fig.~\ref{figure1}. The RPP in Fig.~\ref{figure1}b shows an
$H_{c1}$-pattern with both an elongation in
$[\overline{1}10]$-direction and steps along the hard axes. Both
uniaxial anisotropy components are thus clearly present. From the
length of the diagonals we estimate $\frac{\varepsilon}{M}\sim$ 8
mT. The step height gives $\frac{K_{u1}}{M}\sim$1 mT. From the
rectangle side-ratio we obtain $\delta\sim 8^{\circ}$. $K_{u2}$ is
thus $\sim$15\% of $K_{1}$. $\frac{K_{1}}{M}\sim$100 mT can be
estimated from the asymptotic behavior of the magnetization
towards the hard axes at higher fields and $M\sim 37,000$ A/m is
known from SQUID measurements.

The anisotropy components and $\varepsilon$ can differ from sample
to sample. Neglecting coherent rotation is typically a good model
for the first switching fields $H_{c1}$, whereas $H_{c2}$ is
strongly influenced by magnetization rotation especially along the
hard axes, as can be seen in Fig.~\ref{figure1}b. Sharp switching
events are easily identified by high contrast, whereas gradual
magnetization rotations show up as smooth color transitions.

In conclusion we have shown that RPP compiled from high resolution
IPH measurements constitute a fingerprint of the (Ga,Mn)As
anisotropy. They allow both qualitative and quantitative
statements about the symmetry components of the magnetic
anisotropy and the DW nucleation/propagation energy. The same
technique is equally applicable to any transport phenomena which
produces a response to the orientation of the magnetization, such
as AMR, or TAMR.

We applied the method to several typical MBE layers from our lab
\cite{OtherlabNote} to confirm that biaxial and \textit{both} the
$[\overline{1}10]$ and the [010] uniaxial anisotropy term are
present in all typical compressively strained (Ga,Mn)As layers at 4
K with a relative strength of the order of $K_{1}:K_{u2}:K_{u1}\sim
100:10:1$. Indeed all (Ga,Mn)As layers investigated show both these
uniaxial components, including layers where the [010] component
could not be identified in SQUID measurements. Moreover the
application of our fingerprint method to previously published data
in the literature shows that in all cases where sufficient data is
available, both uniaxial components are present.

The authors thank O. Rival and M. Sawicki for useful discussions
and V. Hock for sample preparation, and acknowledge financial
support from the EU NANOSPIN project (FP6-IST-015728) and the
German DFG (BR1960/2-2) project.

\pagebreak


\begin{thebibliography}{99}

\bibitem{Sawicki} M. Sawicki, F. Matsukura, A. Idziaszek, T. Dietl, G.M. Schott, C. R\"{u}ster, C. Gould, G. Karcewski, G. Schmidt, and L.W. Molenkamp,  {\em Phys. Rev. B.} {\bf 70}, 245325 (2004).

\bibitem{Roukes} H. X. Tang, R.K. Kawakami, D.D. Awschalom and M.L. Roukes {\em Phys. Rev. Lett.} {\bf 90}, 107201 (2003).

\bibitem{ChrisTAMR} C. Gould, C. R\"{u}ster, T. Jungwirth, E. Girgis, G.M. Schott, R. Giraud, K. Brunner, G. Schmidt, and L.W. Molenkamp {\em Phys. Rev. Lett.} {\bf 93}, 117203 (2004).

\bibitem{theory1} T. Dietl, H. Ohno and F. Matsukura, \emph{Phys. Rev B} \textbf{63}, 195205 (2001).

\bibitem{theory2} M. Abolfath, T. Jungwirth, J. Brum and A. MacDonald, \emph{Phys. Rev. B} \textbf{63}, 054418 (2001).

\bibitem{Cowburn} R.P. Cowburn, S.J. Gray, J. Ferr\'{e}, J.A.C. Bland, J. Milltat, {\em J.  Appl. Phys.} {\bf 78}, 7210 (1995)

\bibitem{Jan} a) T.R. McGuire, R.I. Potter, \emph{IEEE Trans. Magn.} \textbf{MAG-11}, 1018 (1975). b) J. P. Jan, in "Solid State Physics" (Eds: F. Seitz, D. Turnbull), Academic Press Inc., New York, 1957.

\bibitem{DWs} Because of the underlying biaxial anisotropy, a $180^\circ$-DW can be seen as two loosly coupled $90^\circ$-DW, thus $\varepsilon_{180^\circ} \sim
2\varepsilon_{90^\circ}$\cite{Cowburn}.

\bibitem{Bland} C. Daboo, R.J. Hicken, D.E.P. Eley, M. Gester, S.J. Gray, A.J.R. Ives, and J.A.C. Bland, \emph{J. Appl. Phys.} \textbf{75},
5586 (1994)

\bibitem{OtherlabNote} Preliminary measurements on samples obtained
from other labs have yielded similar results. Final results on
these will be published elsewhere.


\end{thebibliography}
\end{document}